\documentclass{article}
\usepackage{ijcai20}

\usepackage{times}
\usepackage{soul}
\usepackage{url}
\usepackage[hidelinks]{hyperref}
\usepackage[utf8]{inputenc}
\usepackage[small]{caption}
\usepackage{graphicx}
\usepackage{amsmath}
\usepackage{amsthm}
\usepackage{amssymb}
\usepackage{booktabs}
\usepackage{algorithm}
\usepackage{algorithmic}
\usepackage{amsmath}
\usepackage{xcolor}
\title{Residual Squeeze-and-Excitation Network for Fast Image Deraining}

\author{
Jun Fu$^1$
\and
Jianfeng Xu$^2$\and
Kazuyuki Tasaka$^2$ \And
Zhibo Chen$^1$
\affiliations
$^1$University of Science and Technology of China\\
$^2$KDDI Research, Inc.\\
\emails
fujun@mail.ustc.edu.cn,
\{ji-xu,ka-tasaka\}@kddi-research.jp,
chenzhibo@ustc.edu.cn
}

\begin{document}

\maketitle

\begin{abstract}
Image deraining is an important image processing task as rain streaks not only severely degrade the visual quality of images but also significantly affect the performance of high-level vision tasks. Traditional methods progressively remove rain streaks via different recurrent neural networks. However, these methods fail to yield plausible rain-free images in an efficient manner. In this paper, we propose a residual squeeze-and-excitation network called RSEN for fast image deraining as well as superior deraining performance compared with state-of-the-art approaches. Specifically, RSEN adopts a lightweight encoder-decoder architecture to conduct rain removal in one stage. Besides, both encoder and decoder adopt a novel residual squeeze-and-excitation block as the core of feature extraction, which contains a residual block for producing hierarchical features, followed by a squeeze-and-excitation block for channel-wisely enhancing the resulted hierarchical features. Experimental results demonstrate that our method can not only considerably reduce the computational complexity but also significantly improve the deraining performance compared with state-of-the-art methods. 
\end{abstract}

\section{Introduction}
\begin{figure}[th]
	\centering
	\includegraphics[height=7cm,width=9cm]{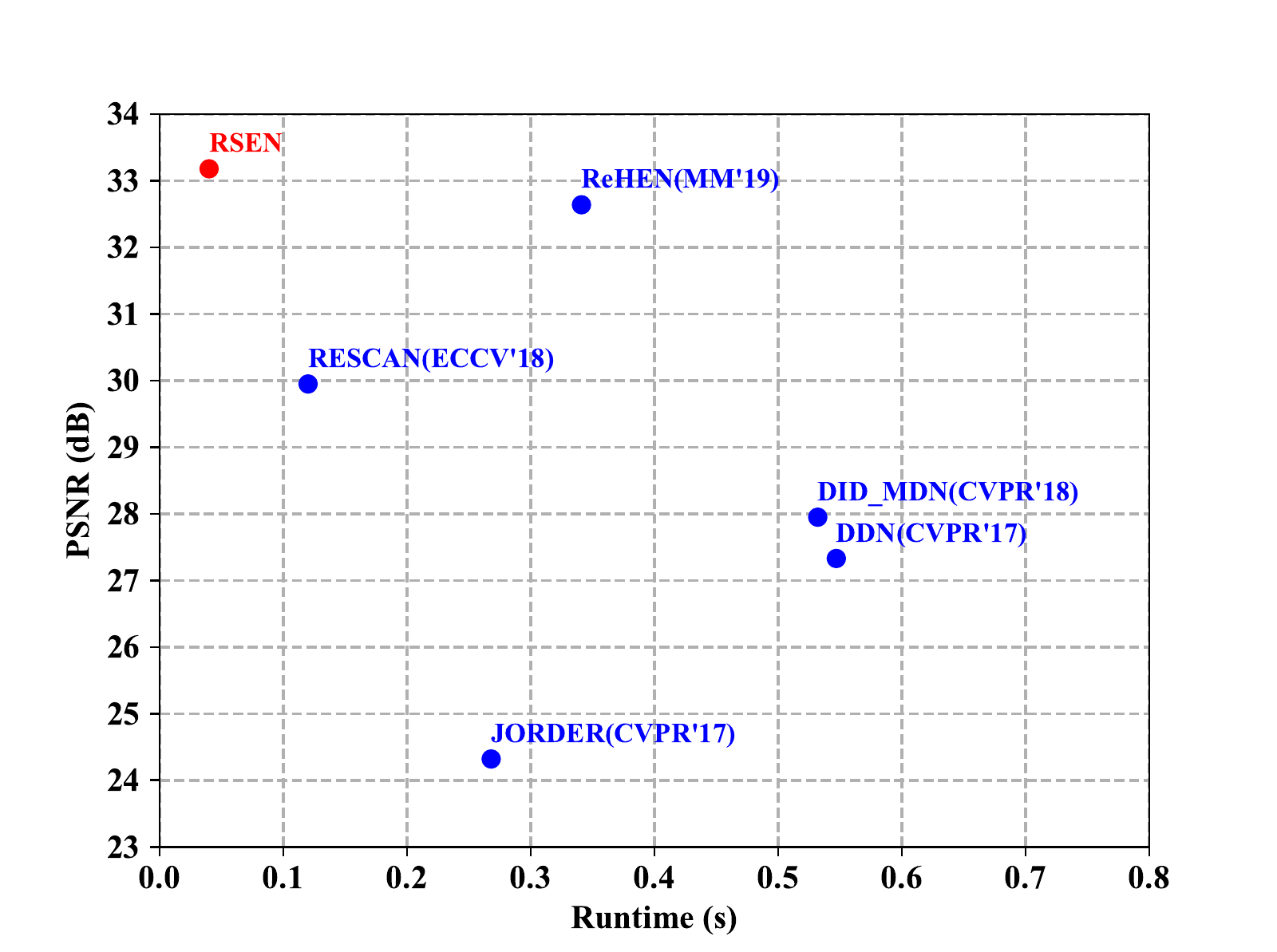}
	\caption {The PSNR-Runtime trade-off plot on the Rain1200 dataset. Compared to five state-of-the-art competitors including DDN, DID\_MDN, JORDER, RESCAN, and ReHEN, RSEN achieves superior quality and lower computational complexity.}
	\label{fig:tradeoff}
\end{figure}
Outdoor images taken on rainy days often contain various rain streaks. These rain streaks not only cause noticeable degradation in scene visibility but also significantly impair the performance of advanced visual tasks, such as pedestrian detection \cite{liu2019high}, object tracking \cite{redmon2018yolov3}, and autonomous vehicles \cite{zang2019impact}. Therefore, it is important and necessary to develop image deraining algorithms. 

In general, a rainy image (I) can be regarded as a linear combination of a clean background (B) and a rain streak layer (R):
\begin{equation}
    \text{I} = \text{B} + \text{R}.
    \label{eq: rain model}
\end{equation}
Image deraining aims to recover the clean background from the observed rainy image. However, due to lacking information on the rain streak layer, rain removal is a serious ill-posed problem.

In the past few decades, image deraining has received considerable attention from industry and academia. Existing methods can be divided into two categories, including model-driven methods and data-driven methods. Model-driven methods can be further divided  into filter based ones \cite{xu2012removing,zheng2013single,ding2016single,kim2013single} and prior based ones \cite{luo2015removing,li2016rain,chang2017transformed}. Considering rain removal as a task of signal filtering, filter based ones utilize physical properties of rain streaks and edge-preserving filters to obtain rain-free images. However, prior based ones formulate rain removal as an optimization problem and utilize various handcrafted image priors to regularize the solution space. Different from model-driven methods, data-driven methods regard rain removal as a task of learning a non-linear function mapping the observed rainy image into the clean background. Motivated by the unprecedented success of deep learning, they model the mapping function with various convolution neural networks (CNNs). Most of them progressively remove rain streaks via different recurrent neural networks \cite{li2018recurrent,ren2019progressive,yang2019single}. Additionally, adversarial learning-based methods \cite{zhang2019image} are proposed to prevent derained images from the blur artifact.

However, existing methods suffer from two key limitations despite achieving deraining performance boost. On the one hand, model-driven methods tend to leave some rain streaks or introduce the blur artifact in derained images. This is because physical properties of rain streaks and handcrafted image priors are easily violated on real-world examples where rain streaks are far more complex than modeled. Furthermore, these methods involve heuristic parameter tuning and expensive computation. Data-driven methods, on the other hand, fail to yield plausible rain-free images in an efficient manner (as shown in Fig. \ref{fig:tradeoff}). Although some lightweight methods \cite{fan2018residual,fu2019lightweight} have been proposed to improve the computational efficiency, they result in a significant decrease in the deraining performance. 

Motivated by addressing above two issues, we propose a residual squeeze-and-excitation network called RSEN for fast image deraining as well as superior deraining performance compared with state-of-the-art approaches. Unlike prevalent recurrent networks reusing flat network structures, RSEN adopts a lightweight encoder-decoder architecture to conduct rain removal in one stage. Besides, both encoder and decoder adopt a novel residual squeeze-and-excitation block as the core of feature extraction, which contains a residual block for producing hierarchical features, followed by a squeeze-and-excitation block for channel-wisely enhancing the resulted hierarchical features.

Main contributions of this paper are listed as follows:
\begin{itemize}
\item We propose a novel residual squeeze-and-excitation network called RSEN for image rain removal. It adopts a lightweight encoder-decoder architecture and is capable to effectively remove rain streaks while well preserving texture details.
\item We propose to incorporate a residual squeeze-and-excitation block in our network, which can not only generate channel-wisely enhanced hierarchical features but also well benefit gradient propagation.
\item Experimental results show that our proposed method can not only considerably reduce the computational complexity but also significantly improve the deraining performance compared with state-of-the-art methods. 
\end{itemize}

The remainder of this paper is organized as follows. Section 2 discusses the related works. Section 3 introduces the proposed network. Then, performance evaluation and comparison are presented in Section 4. Finally, Section 5 concludes the paper and discusses future work.

\begin{figure*}[h]
	\centering
	\includegraphics[height=6.5cm, width=18cm]{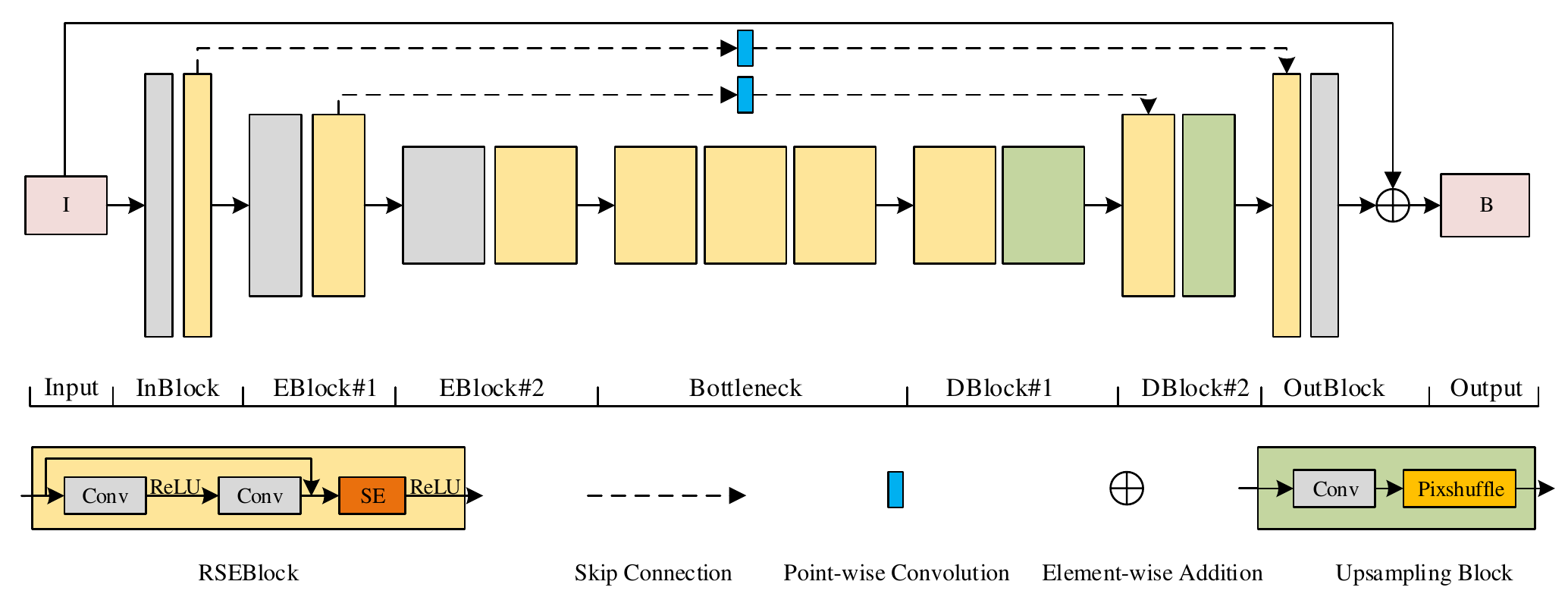}
	\caption {The whole framework of the proposed residual squeeze-and-excitation network. I, B, Conv, and SE represent the input rainy image, the clean background, the conventional convolution  and the squeeze-and-excitation block, respectively.}
	\label{fig:framework}
\end{figure*}

\section{Related Work}
\subsection{Image Deraining}
Image deraining, a highly ill-posed problem, has drawn increasingly more attention from industry and academia over the past few decades. Existing works can be categorized into two classes, i.e., model-driven methods and data-driven methods. 

Model-driven methods can be further divided into filter based ones and prior based ones. Considering rain removal as a task of signal filtering, filter based ones utilize physical properties of rain streaks and edge-preserving filters to obtain rain-free images. Specifically, \cite{xu2012removing} utilize guided filter \cite{he2010guided} to remove rain streaks based on the chromatic and brightness property of rain streaks. \cite{zheng2013single}, \cite{kim2013single}, and \cite{ding2016single} boost the deraining performance via multi-guided filter, non-local means filtering, and guided $L_0$ smoothing filter, respectively. However, prior based ones formulate rain removal as an optimization problem and employ various handcrafted image priors to regularize the solution space. These image priors include sparse-coding prior \cite{luo2015removing},  Gaussian prior \cite{li2016rain} and low-rank prior \cite{chang2017transformed}. 

Different from model-driven methods, data-driven approaches employ various convolution neural networks to automatically learn a non-linear mapping function between the rainy image and the rain-free image from data. More specifically, \cite{fu2017clearing} design a shallow convolution neural network to address rain removal and improve the deraining performance by a deeper network \cite{fu2017removing}. \cite{yang2017deep} present a multi-task framework, which simultaneously deals with rain detection and rain removal. Unlike one-stage deraining methods, \cite{li2018recurrent} remove rain streaks stage by stage via a recurrent squeeze-and-excitation context aggregation network.  Recently, \cite{wang2019spatial} present a spatial attentive single-image deraining approach and \cite{yang2019single} introduce a recurrent hierarchy enhancement network. To preserve more texture details of derained images, adversarial learning-based methods \cite{zhang2019image} are proposed. In addition, some lightweight methods \cite{fu2019lightweight,fan2018residual} are dedicated to reduce the computational complexity. 

However, existing methods suffer from two key limitations despite achieving deraining performance boost. On the one hand, model-driven methods suffer from under-/over-deraining on real-world examples where rain streaks are far more complex than modeled. On the other hand, existing neural networks fail to produce plausible rain-free images in an efficient manner due to the complex framework. Some lightweight networks attempt to improve  computational efficiency but at the cost of obvious performance degradation.   

\subsection{Convolution Neural Network}
Recent years have witnessed the convolution neural network goes increasingly deeper (e.g., VGGNet \cite{simonyan2014very}).  Deep neural networks typically are superior to shallow networks while meeting more challenges in convergence and generalization. To address this problem, \cite{he2016deep} first propose residual learning, which can benefit gradient propagation and accelerate convergence. In addition,  diverse attention mechanisms are proposed to improve the capability of neural networks, such as channel-wise attention \cite{hu2018squeeze}, self-attention mechanism \cite{vaswani2017attention} and spatial attention \cite{woo2018cbam}. 

\section{Proposed Method}
\subsection{Problem Formulation}
Considering rain streak accumulation and overlapping,  \cite{yang2017deep} proposes a rain model as follows:
\begin{equation}
    I = \alpha (B + \sum_{t=1}^{n} S_t M) + (1-\alpha) A,
    \label{eq:rain model 1}
\end{equation}
where each $S_t$ denotes a layer of rain streaks that have the same direction, $n$ denotes the total number of rain streak layers, $M$ records the locations of $S_t$, $A$ represents the global atmospheric light, and $\alpha$ is the atmospheric transmission.  

Unlike the above model requiring rain detection, \cite{li2018recurrent} present a simpler model, i.e., dividing the captured rainy scene into the combination of several rain streak layers and a rain-free background. Thus, the rain model can be reformulated as follows:
\begin{equation}
    I = B + \sum_{t=1}^{n} R_t,
\end{equation}
where $R_t$ and $n$ represent the $t$-th rain streak layer that consists of one kind of rain streaks and the maximum number of rain streak layers, respectively.  

In this paper, we further simplify the rain model as follows:
\begin{equation}
    \text{I} = \text{B} + \text{R}.
    \label{eq:rain model 3}
\end{equation}
As Eq. \ref{eq:rain model 3} implied, we can obtain the rain-free background by subtracting rain streaks $R$ from the rainy image $I$. Therefore, we formulate the rain removal as a task of learning a non-linear mapping function between the rainy image and rain streaks. Inspired by the recent success of deep learning, we propose a residual squeeze-and-excitation network call RSEN to model the non-linear mapping function. As illustrated in Fig. \ref{fig:framework}, RSEN adopts a commonly used encoder-decoder architecture. The detailed design of this architecture is introduced and explained next. 

\subsection{Residual Squeeze-and-Excitation Network}
The encoder-decoder network typically adopts a symmetric convolution neural network architecture composed of an encoder and a decoder. The encoder transforms the input data into feature maps with smaller spatial sizes and more channels while the decoder transforms the resulted feature maps back to the shape of the input. In addition, skip connections are widely used in this architecture because they can aggregate features at multiple levels and accelerate convergence. 

Compared with flat architectures, the encoder-decoder network has shown its superiority in many low-level vision tasks, such as image deblurring \cite{tao2018scale} and image inpainting \cite{nazeri2019edgeconnect}. However, to accommodate specific tasks, such a network needs to be carefully designed.

For the task of image deraining, we take the following three key aspects into account in our design. First, the receptive field needs to be large enough to handle heavy rain streaks. To this end, a naive approach is to stack more levels or adding more convolution layers at each level for encoder/decoder modules. However, this strategy will result in a sharp increase in computational complexity and parametric size. Furthermore, such a deep neural architecture generally suffers from a low speed of convergence. Second, aside from global skip connections, local skip connections are also beneficial to gradient propagation and accelerate convergence. Third, according to the experimental results of \cite{li2018recurrent} and \cite{yang2017deep}, the channel-wise attention mechanism is a promising alternative to improve the deraining performance.

Based on the aforementioned analysis, we adapt the encoder-decoder network into our task as follows. First, we propose a residual squeeze-and-excitation block (RSEBlock), which contains a residual block for producing hierarchical features via local skip connections, followed by a squeeze-and-excitation block for channel-wisely enhancing the resulted hierarchical features. It is worth noting that we remove the batch normalization \cite{ioffe2015batch} from the original residual blocks in ResNet \cite{he2016deep} according to \cite{tao2018scale} and our experimental results. Second, both encoder and decoder employ the RSEBlock as their core of feature extraction instead of conventional convolution layers. Third, considering that the spatial size of the middle feature map needs to be large enough to keep sufficient spatial information for reconstruction, we only stack two levels in the encoder and the decoder. Finally, to enlarge the size of the receptive field, we stack three RSEBlocks after the encoder. 

The proposed network can be mathematically expressed as

\begin{equation}
\begin{aligned}
    f &= Net_E(I;\theta_E), \\
    h &= Net_B(f; \theta_{B}),\\
    \hat{B} &= I - Net_D(h;\theta_D),
\end{aligned}
\end{equation}
where $Net_E$, $Net_B$ and $Net_D$ are encoder, bottleneck, and decoder CNNs with parameters $\theta_E$, $\theta_B$ and $\theta_D$. $f$ and $h$ are the intermediate feature maps. $\hat{B}$ is the recovered rain-free image.

The implementation details of our proposed RSEN are specified as follows. In our proposed architecture, there are 1 InBlock, 2 EBlocks, followed by 1 Bottleneck, 2 DBlocks, and 1 OutBlock, as illustrated in Fig. \ref{fig:framework}. InBlock composed of a convolution layer and a RSEBlock transforms the input 3-channel rainy image into a 64-channel feature map. Each EBlock adopts the same structure as InBlock while doubling the number of kernels in the previous layer and downsampling feature maps by half. DBlocks is symmetric to EBlock. It is designed to double the spatial size of feature maps and halve channels, composed of a  RSEBlock and a upsampling block. In the upsampling block, a point-wise convolution is used to increase the channel dimension of the input 4 times, followed by a pix-shuffle layer \cite{shi2016real} whose scale factor is set to 2. Additionally, 2 point-wise convolutions are designed for skip connections. The Bottleneck contains 3 RESBlocks that have the same number of channels. OutBlock takes previous feature maps as input and generates estimated rain streaks. All squeeze-and-excitation blocks squeeze the channel dimension of the input feature map to 6. The stride size for the convolution layer in EBlocks is 2, while all others are 1. Rectified Linear Units (ReLU) are used as the activation function for all layers, and all kernel sizes are set to 3. 

\subsection{Loss Function}
The loss function is defined as the mean square error (MSE) between the derained image and its corresponding groundtruth, which can be formulated as follows:
\begin{equation}
    L = \frac{1}{2HWC} \sum_i \sum_j \sum_k  \lvert\lvert \hat{B}_{i,j,k} - B_{i,j,k} \rvert\rvert^2_2,
\end{equation}
where $H$, $W$, and $C$ represent the height, width, and channel number of the rain-free image. $\hat{B}$ and $B$ denote the restored rain-free image and the groundtruth, respectively.

\begin{table}[t]
\caption{Details of synthetic and real-world datasets. Values in each column of the training set and testing set indicate the
number of rain-free/rainy image pairs with the exception of the real-world
set with rainy images only.}
    \small
    \centering
    \begin{tabular}{l|r|r|c}
        \hline
        Datasets & Training Set & Testing Set & Label \\
        \hline
        \hline
        Rain100H & 1800 & 100 & rain mask/rain map \\
        Rain800 & 700 & 100 & - \\
        Rain1200 & 12000 & 1200 & rain mask/rain map \\
        Rain1400 & 12600 & 1400 & - \\     
        \hline
        Real-world set & - & 13 & - \\
        \hline
    \end{tabular}
    \label{tab:datasets}
\end{table} 
 
\section{Experiments}
\subsection{Dataset and Evaluation Metrics}
We choose four synthetic datasets and a real-world dataset for the deraining experiment. The synthetic datasets include Rain100H, Rain800, Rain1200 and Rain1400. Rain100H is synthesized by \cite{yang2017deep}. Rain800 is obtained from \cite{zhang2019image}. Rain1200 is provided by \cite{zhang2018density} while Rain1400 comes from \cite{fu2017removing}. The real-world dataset is constructed by \cite{yang2017deep}. Rainy images in these datasets are diverse in terms of content as well as the type of rain streaks. More details of these datasets are listed in Table \ref{tab:datasets}. 

Deraining performances on the synthetic datasets are evaluated in terms of peak signal-to-noise (PSNR) \cite{huynh2008scope} and structural similarity (SSIM) \cite{wang2004image}. Due to lacking the groundtruth of real-world rainy images, performances of different methods on the real-world dataset are evaluated visually. We compare RSEN with state-of-the-art methods including DSC \cite{luo2015removing}, LP \cite{li2016rain}, JCAS \cite{gu2017joint}, DDN \cite{fu2017removing}, JORDER \cite{yang2017deep}, DID-MDN \cite{zhang2018density}, DualCNN \cite{pan2018learning}, RESCAN \cite{li2018recurrent}, ID\_CGAN \cite{zhang2019image}, ReHEN \cite{yang2019single}, SPANET \cite{wang2019spatial}, and PReNET \cite{ren2019progressive}. 

\subsection{Training Details}
We implement our model using the PyTorch \cite{paszke2017automatic} framework. During training, 4 rain-free/rainy patch pairs with a size of 256 $\times$ 256 are randomly generated from input image pairs per iteration. All trainable variables of the proposed RSEN are initialized by the default initializer and optimized via the Adam optimizer \cite{kingma2014adam}. $\beta_1$, $\beta_2$ and $\epsilon$ of the Adam optimizer are set to 0.9, 0.999, and $10^{-8}$ respectively. We train RSEN on a NVIDIA Geforce GTX 1080Ti GPU for 700 epochs. The learning rate is initialized as $10^{-4}$ and decayed in half every 150 epochs. For fair comparison with existing methods, we only use rain-free/rainy image pairs for training without other additional labels and any data augmentation. 

\begin{table*}[]
\centering
\caption{Parametric size, running time, and average PSNR and SSIM values on four synthetic datasets. The value with red bold font denotes ranking the first place in this column while value with blue font is the second place. It is worth noting that both PSNR and SSIM are calculated in the RGB color space.}
\begin{tabular}{lrclcccccccc} \hline
                 & & \multicolumn{2}{c}{Time (s)}    & \multicolumn{2}{c}{Rain100H} & \multicolumn{2}{c}{Rain800} & \multicolumn{2}{c}{Rain1200} & \multicolumn{2}{c}{Rain1400} \\ \cline{3-12}
Methods  &    Params      & \multicolumn{2}{c}{512x512} & PSNR          & SSIM         & PSNR         & SSIM         & PSNR          & SSIM         & PSNR          & SSIM         \\ \hline \hline
Rainy             & - & \multicolumn{2}{c}{-}        & 12.13         & 0.349       & 21.16        & 0.652       & 21.15         & 0.778        & 23.69         & 0.757        \\
DSC(ICCV'15)      & - & \multicolumn{2}{c}{-}        & 15.66         & 0.544        & 18.56        & 0.599        & 21.44         & 0.789        & 22.03         & 0.799        \\
LP(CVPR'16)       & - &\multicolumn{2}{c}{-}        & 14.26         & 0.423        & 22.27        & 0.741        & 22.75         & 0.835        & 25.64         & 0.836        \\
JCAS(ICCV'17)     & - &\multicolumn{2}{c}{-}        & 13.65         & 0.459        & 22.19        & 0.766        & 27.91         & 0.778        & 28.77         & 0.819        \\
DDN(CVPR'17)      & 57,369 &\multicolumn{2}{c}{0.547}        & 24.95         & 0.781        & 21.16        & 0.732        & 27.33         & 0.898        & 27.61         & 0.901        \\
JORDER(CVPR'17)   & 369,792 &\multicolumn{2}{c}{0.268}        & 22.15         & 0.674        & 22.24        & 0.776        & 24.32         & 0.862        & 27.55         & 0.853        \\
DID-MDN(CVPR'18)  & 372,839 &\multicolumn{2}{c}{0.532}        & 17.39         & 0.612        & 21.89        & 0.795        & 27.95         & 0.908        & 27.99         & 0.869        \\
DualCNN(CVPR'18)  & 687,008 &\multicolumn{2}{c}{20.19}        & 14.23         & 0.468        & 24.11        & 0.821        & 23.38         & 0.787        & 24.98         & 0.838        \\
RESCAN(ECCV'18)   & 134,424 &\multicolumn{2}{c}{0.281}        & 26.45         & 0.846        & 24.09        & 0.841        & 29.95         & 0.884        & 28.57         & 0.891        \\
ID\_CGAN(TCSVT'19)     & 817,824 &\multicolumn{2}{c}{0.286}        & 14.16         & 0.607        & 22.73        & 0.817        & 23.32         & 0.803        & 21.93         & 0.784        \\
ReHEN(MM'19)      & 298,263 &\multicolumn{2}{c}{0.181}        & 27.97         & 0.864        & \textcolor{blue}{26.96}         & \textcolor{blue}{0.854}       & \textcolor{blue}{32.64}          & \textcolor{blue}{0.914}        & \textcolor{blue}{31.33}          & \textcolor{blue}{0.918}         \\
SPANet(CVPR'19)   & 283,716 &\multicolumn{2}{c}{2.301}        &     -         &      -       &      -        &  -            & 28.64         & 0.91              &    -           &    -               \\ 
PReNET(CVPR'19)   & 168,693 &\multicolumn{2}{c}{0.461}        & \textcolor{blue}{28.06}        & \textcolor{blue}{0.888}          &      -        &     -         & -         & -             &   30.73            &            0.918         \\ \hline
Ours              & 4,851,373 &\multicolumn{2}{c}{0.040}      & \textcolor{red}{30.86}        & \textcolor{red}{0.902}          & \textcolor{red}{27.58}        &   \textcolor{red}{0.856}      & \textcolor{red}{33.19}         &   \textcolor{red}{0.923}     & \textcolor{red}{31.64}         & \textcolor{red}{0.924}    \\
\hline
\end{tabular}
\label{tab: all}
\end{table*}

\begin{table*}[]
\centering
\caption{Ablation study on different modules of the proposed RSEN. The CoarseNet denotes the architecture without any enhancement tricks.}
\begin{tabular}{l ll ll ll ll} \hline
           & \multicolumn{2}{c}{Rain100H} & \multicolumn{2}{c}{Rain800} & \multicolumn{2}{c}{Rain1200} & \multicolumn{2}{c}{Rain1400} \\ \cline{2-9}
Method              & PSNR            & SSIM          & PSNR       & SSIM          & PSNR       & SSIM          & PSNR         & SSIM \\ \hline \hline
CoarseNet          &   28.28          & 0.822         & 21.42      & 0.647            & 32.64   & 0.908  &  31.36 & 0.917       \\
CoarseNet + Skip    &   30.02          & 0.888         & 26.47      & 0.845            & 32.88 & 0.919 & 31.45 & 0.922           \\
CoarseNet + Skip  + RES     & 30.86     & 0.902     & 26.90   & 0.854   & 33.11 & 0.923  & 31.54 & 0.924     \\
CoarseNet + Skip  + RES + SE  &  30.30 & 0.893     &   27.58         &  0.856     & 33.19         &   0.923    &31.64         & 0.924    \\
\hline
\end{tabular}
\label{tab:ablation}
\end{table*}

\begin{figure}[h]
	\centering
	\includegraphics[height=5.8cm,width=6.8cm]{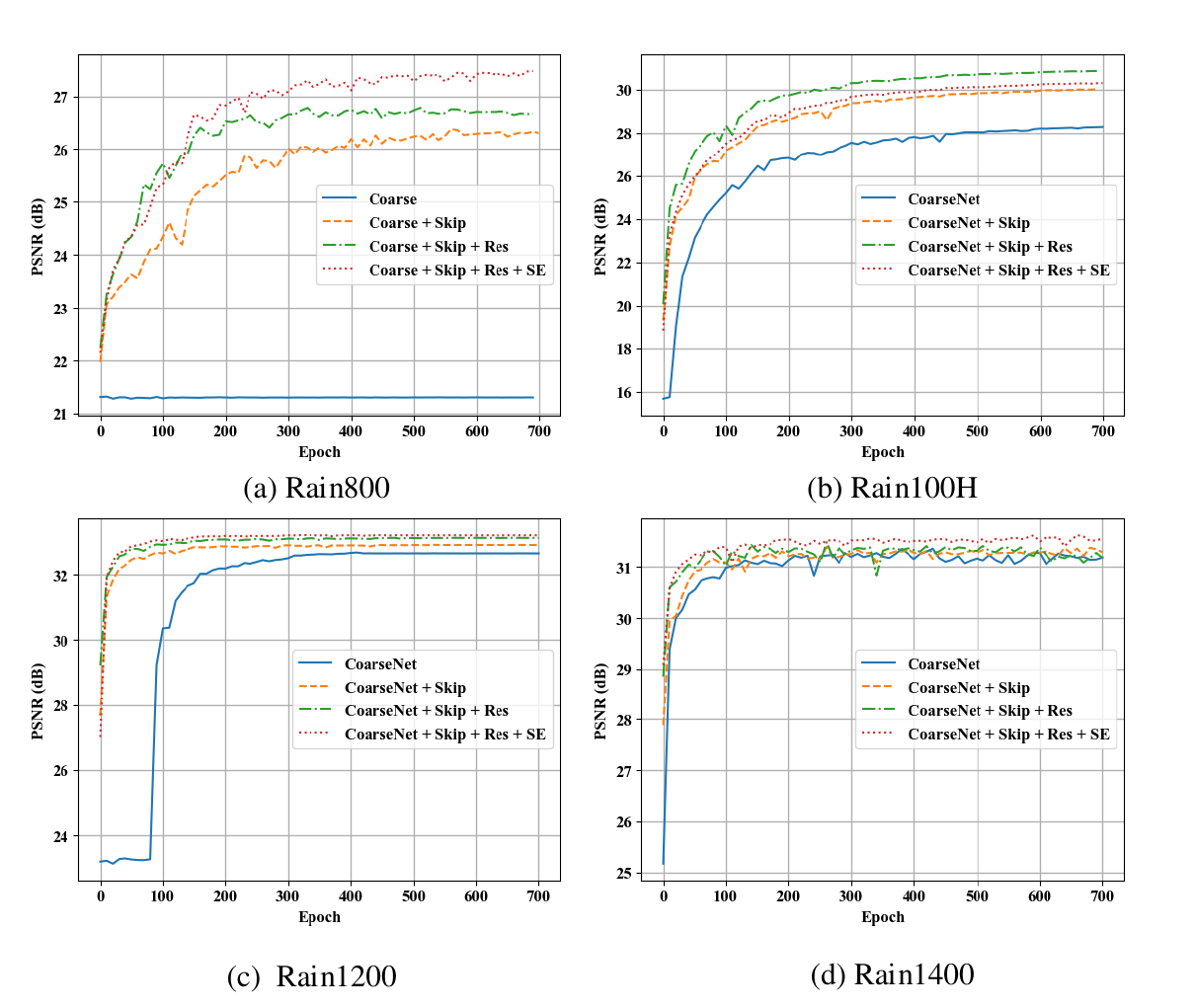}
	\caption {Training convergence analysis on PSNR of RSEN with different modules.  }
	\label{fig:ablation}
\end{figure}

\begin{figure*}[h]
	\centering
	\includegraphics[height=7.5 cm,width=16cm]{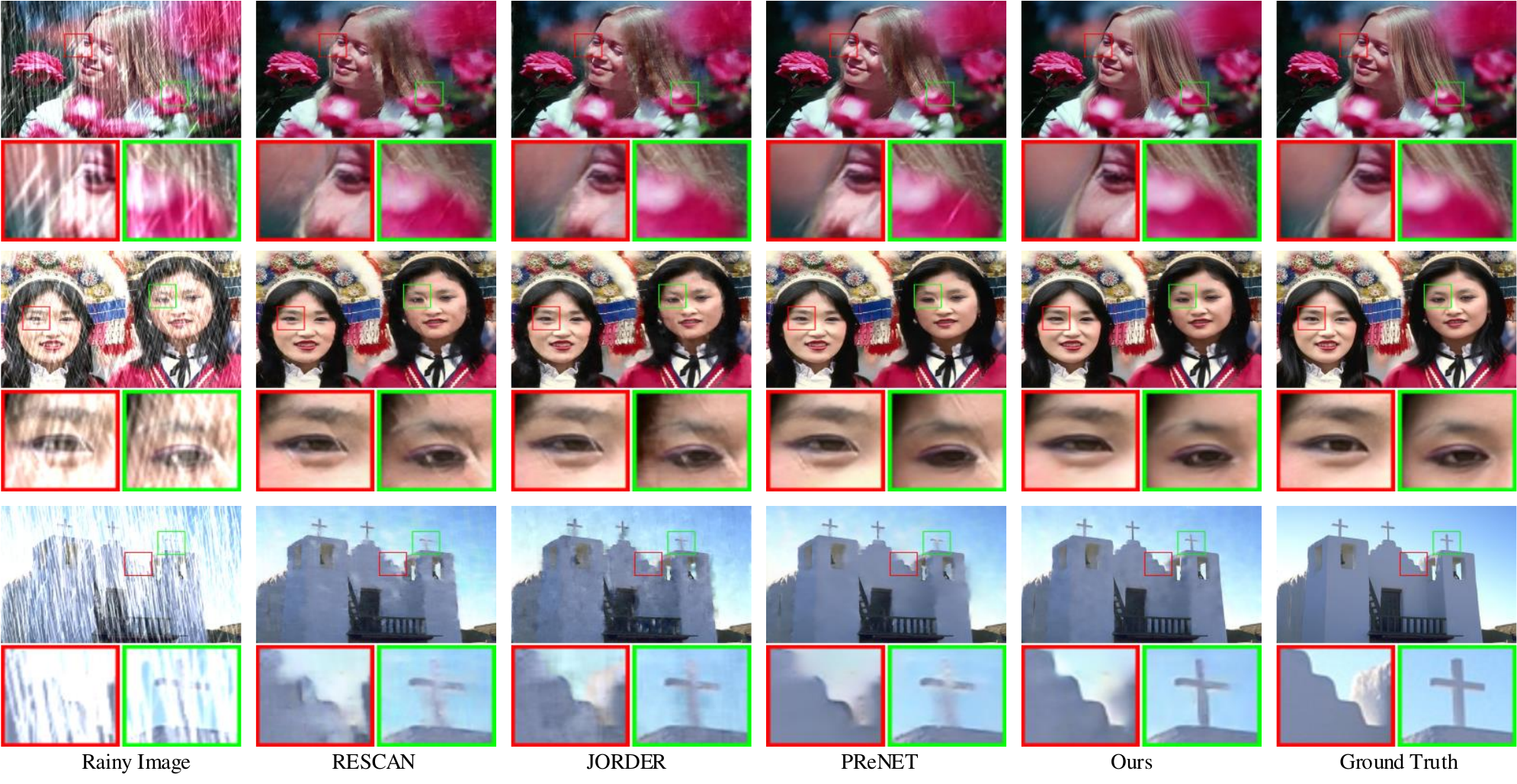}
	\caption {Derained results of RESCAN, JORDER, PReNET, and the proposed method on synthetic rainy images.}
	\label{fig:synthetic}
\end{figure*}

\begin{figure*}[!htb]
	\centering
	\includegraphics[height=7.5 cm,width=16cm]{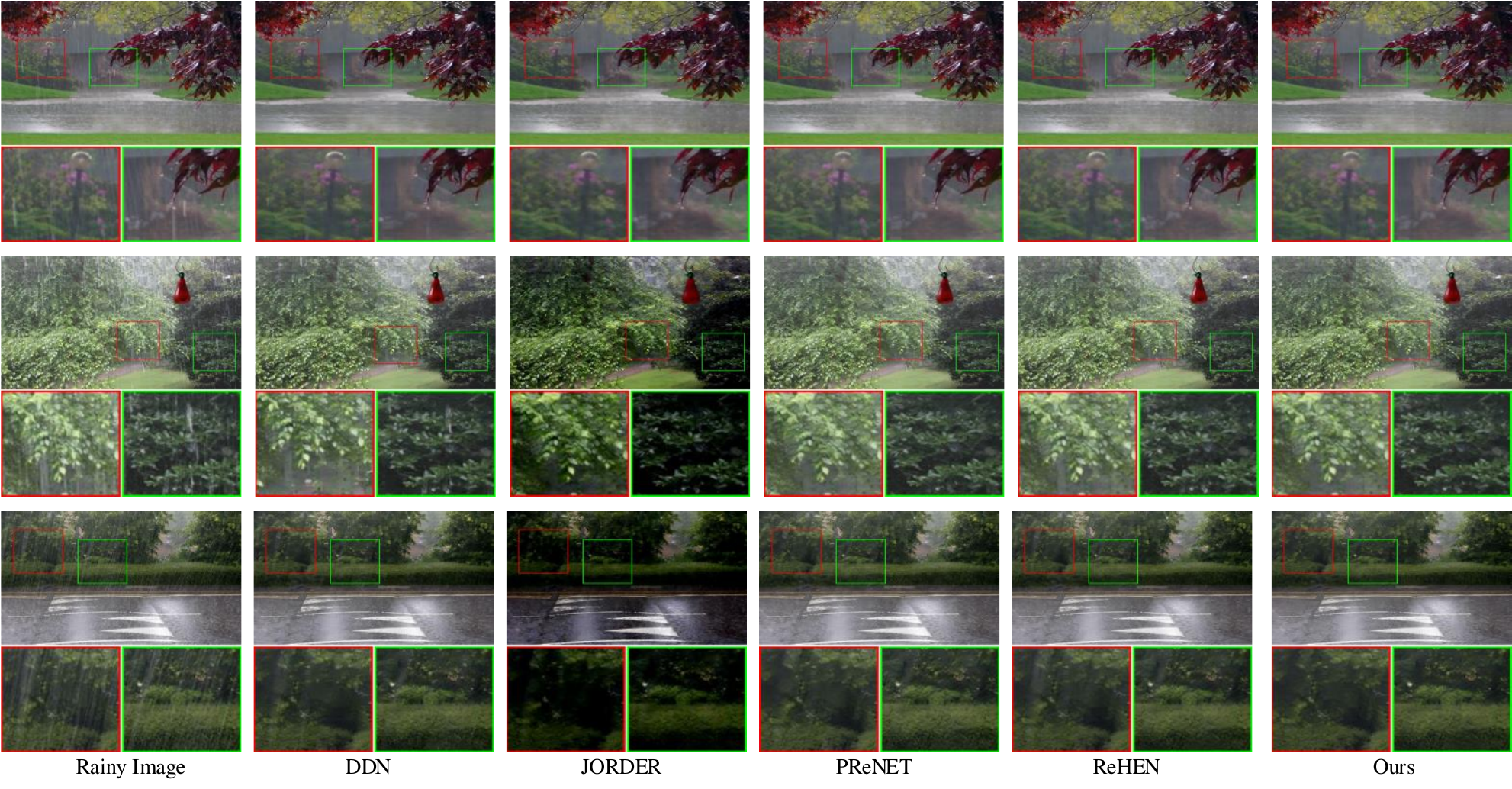}
	\caption {Derained results of DDN, JORDER, PReNET, ReHEN, and the proposed method on real-world rainy images.}
	\label{fig:real-world}
\end{figure*}

\subsection{Results on Synthetic Dataset}
Table \ref{tab: all} presents quantitative results of different methods on four synthetic datasets. We can observe that the proposed method achieves substantial improvements over state-of-the-art approaches in terms of both PSNR and SSIM. Specifically, the proposed method obtains significant improvements by PSNR of 2.80 dB, 0.62 dB, 0.55 dB, 0.31 dB and SSIM of 0.014, 0.002, 0.09, 0.06 on Rain100H, Rain800, Rain1200, and Rain1400, respectively. Average running time for a rainy image with a size of 512 $\times$ 512 and the parametric size of each method are also chosen for comparison. It can be observed that the proposed method significantly reduces the computational complexity in spite of using relatively more parameters compared with other methods. For instance, RSEN reaches around 4, 10 and 50 times faster than ReHEN, PReNET, and SPANet, respectively. This confirms the superiority of our proposed architecture. 

Fig. \ref{fig:synthetic} visually compares the proposed method with three recent state-of-the-art methods. We can observe that the proposed method achieves the best visual quality. Specifically, compared with other methods, our proposed method avoids leaving obvious artifacts (e.g., rain streaks and the blur artifact) in derained images as well as preserving more structural information.  

\subsection{Results on Real-World Dataset}
To assess the generalization of the proposed method, we also evaluate the proposed method on a real-world dataset. For fair comparison, all the methods employ the pre-trained model trained on the Rain100H dataset to remove rain streaks from real-world rainy images. As demonstrated in Fig. \ref{fig:real-world}, the proposed method produces more natural and pleasant derained images compared with four state-of-the-art methods. Specifically, in our experiment, DDN fails to remove rain streaks from real-world rainy images in most cases while JORDER is prone to dim and blur the details of the derained results. Compared with PReNET and ReHEN, the proposed method can more effectively remove rain streaks from real-world rainy images and restore more texture details. 

\subsection{Ablation Study}
Table \ref{tab:ablation} and Fig. \ref{fig:ablation} show the ablation investigation on the effects of the skip connection (SKip), the residual connection (RES), and the squeeze-and-excitation block (SE). As can be observed, both Skip and RES are beneficial for convergence and boosting deraining performance. Specifically, the Skip brings gains by PSNR of 1.74 dB, 5.02 dB, 0.24 dB and 0.09 dB on Rain100H, Rain800, Rain1200, and Rain1400, respectively, while the RES obtains 0.84 dB, 0.43 dB, 0.23 dB, and 0.09 dB gains. The SE is useful for most datasets except for Rain100H. This may be because synthesized rainy images with five streak directions in Rain100H are more complex than other datasets with one streak direction. Hence, we disable the SE in all the RSEBlocks when performing rain removal on Rain100H.

\section{Conclusion}
In this paper, we propose a novel residual squeeze-and-excitation network called RSEN for fast image deraining as well as superior deraining performance compared with existing approaches. Specifically, RSEN adopts a lightweight encoder-decoder architecture to conduct rain removal in one stage. Besides, both encoder and decoder adopt a novel residual squeeze-and-excitation block as the core of feature extraction, where a residual block and a squeeze-and-excitation block are used to produce and channel-wisely enhance hierarchical features, respectively. Experimental results demonstrate that our method can not only considerably reduce the computational complexity but also achieve significant improvements in the deraining performance compared with state-of-the-art methods. In the future, we plan to extend our work to deal with video deraining, e.g., incorporating an extra module in our network to capture temporal information. 

\scriptsize
\bibliographystyle{named}
\bibliography{ijcai20}

\end{document}